# Hyperdisorder in tumor growth

Arturo Tozzi (corresponding author)
ASL Napoli 1 Centro, Distretto 27, Naples, Italy
Via Comunale del Principe 13/a 80145
tozziarturo@libero.it

ABSTRACT

Tumor growth is constrained by spatial, mechanical and metabolic factors whose alignment progressively breaks down across cellular, mesoscopic and tissue scales as tumors expand. We hypothesize that this misalignment could drive growing tumors toward a distinct architectural regime, termed hyperdisorder. Hyperdisorder is not defined by increased heterogeneity alone, but by the coexistence of elevated disorder across scales and spatial nonstationarity within the same tumor. Unlike ordinary randomness, where independent fluctuations diminish under spatial averaging, here disorder persists, reorganizes or even amplifies with increasing observation scale, preventing convergence toward a stable architectural description. Using hematoxylin and eosin stained whole-slide images of gastric cancer from The Cancer Genome Atlas, we quantify tumor architecture with tile-based metrics capturing complementary aspects of organization: texture entropy, microstructural fragmentation orientation isotropy and multiscale entropy variation. These measures are combined into a standardized hyperdisorder index, enabling unsupervised comparison across spatial regions. We find that architectural disruption is unevenly distributed and partially decoupled across scales within individual slides, consistent with growth-driven multiscale incoherence rather than uniform stochastic variability. Testable consequences include anomalous scaling of heterogeneity with sampling size, failure of coarse graining to converge and systematic differences between tumor cores and invasive fronts. In diagnostic and clinical contexts, this perspective clarifies when measurements derived from limited tissue samples are representative of the tumor as a whole and when they are instead dominated by scale-dependent and location-specific architectural effects. In therapeutic settings, it enables quantitative assessment of pre-existing architectural differences among spatially distinct tumor regions prior to treatment.

INTRODUCTION

Histopathological assessment of cancer typically relies on tissue architecture evaluated at a single spatial scale or on quantitative descriptors computed locally. Grading systems, pattern recognition approaches and many image-based metrics implicitly assume that architectural disruption is a local phenomenon and that variability smooths out as larger tissue regions are considered (Shekarian et al. 2017; Bai et al. 2020; Man and Jenkins 2022; Wang et al. 2023). These assumptions limit the ability of current methods to capture how heterogeneity is organized across spatial scales during tumor growth. Common measures often conflate structural complexity with randomness, interpreting disorder as a uniform increase in variability (Ren et al. 2018; Bagaev et al. 2021; Leong et al. 2022). However, architectural disruption need not arise from independent random perturbations. Instead, disorder may emerge from growth dynamics, producing patterns that are distributed, partially coupled and maintained across multiple spatial scales. This motivates the introduction of a distinct architectural regime, termed hyperdisorder, that cannot be reduced to conventional notions of heterogeneity.
Hyperdisorder describes a condition in which tissue organization becomes progressively incoherent across spatial scales as a tumor grows. Unlike ordinary randomness, where fluctuations diminish predictably under spatial averaging, hyperdisorder denotes a regime in which disorder persists, reorganizes or increases with scale (Hayashi 2023; Mittal 2025). It is characterized by elevated disorder at multiple scales, partial decoupling between modes of architectural variation and strong spatial nonstationarity within the same specimen (Wang et al. 2025). Cellular-scale irregularities, mesoscopic fragmentation and tissue-scale patterns fail to align into a single degraded architecture and do not converge toward a stable description (Ross et al. 2025). Hyperdisorder is defined operationally through measurable image-derived quantities and their scaling behavior. We apply this framework to gastric cancer histology to test whether tumor architecture exhibits signatures of growth-driven multiscale incoherence and whether these differ from those expected under ordinary stochastic heterogeneity.

We proceed as follows. We first describe the image data and quantitative measures used to characterize tissue architecture and introduce a hyperdisorder index with its associated scaling analyses. We then present results from the gastric cancer case study and derive quantitative, testable predictions. Finally, we discuss the implications for interpreting architectural variability in growing tumors, including diagnostic, clinical and therapeutic relevance.

METHODS

We describe here the data sources, preprocessing steps, quantitative measures, mathematical formulations and analytical procedures to operationalize hyperdisorder in tumor histology.

**Histological dataset.** We analyzed publicly available histological data from the stomach adenocarcinoma cohort of The Cancer Genome Atlas (TGCA). Specifically, we used the TCGA stomach histology image collection distributed through the Kaggle repository curated by Ahmed Abo Enaba. This dataset consists of hematoxylin and eosin stained whole-slide images that were pre-tiled into fixed-size, non-overlapping image patches extracted from diagnostic tumor slides. At the time of analysis (January 2026), the repository contained 13 527 tumor tiles derived from 38 patients, each tile corresponding to a square tissue region sampled at a fixed pixel resolution.
To reduce patient-level imbalance, we constructed a stratified subsample with a maximum of 200 tiles per patient, yielding an initial set of 5 702 tiles. We then applied a quality-based filtering step to remove tiles dominated by background or empty regions. After filtering, 899 tiles from 37 patients remained and were retained for quantitative analysis (Fig. A). All analyses and results reported in this manuscript are based exclusively on this final tile set.

**Preprocessing and tissue masking.** Each image tile was first resized to a common spatial resolution to standardize subsequent computations. Let $I(\mathbf{r})$ denote the color image defined on pixel coordinates $\mathbf{r} = (x, y)$. Tiles were isotropically resized to $128 \times 128$ pixels using area interpolation to preserve average intensity statistics. The resized image was converted from RGB to both grayscale and HSV color spaces. The grayscale image $G(\mathbf{r})$ was used for intensity-based and gradient-based measures, while the saturation channel $S(\mathbf{r})$ from HSV was used to discriminate tissue from background.
A binary tissue mask $M(\mathbf{r})$ was constructed according to

$$M(\mathbf{r}) = \begin{cases} 1, & G(\mathbf{r}) < \tau_g \text{ and } S(\mathbf{r}) > \tau_s, \\ 0, & \text{otherwise}, \end{cases}$$

with fixed thresholds $\tau_g = 230$ and $\tau_s = 18$ in 8-bit intensity units. The tissue fraction was defined as

$$f_{\text{tissue}} = \frac{1}{N} \sum_{\mathbf{r}} M(\mathbf{r}),$$

where $N$ is the total number of pixels. Tiles with $f_{\text{tissue}} < 0.10$ were excluded to avoid spurious measurements driven by background noise. All subsequent computations were restricted to pixels with $M(\mathbf{r}) = 1$, ensuring that quantitative measures reflect tissue architecture rather than slide artifacts.

**Local intensity entropy as a disorder measure.** To capture variability in staining and microtexture, we computed the Shannon entropy of grayscale intensities within tissue regions. Let $g(\mathbf{r})$ denote grayscale intensity values restricted to tissue pixels. The empirical histogram $h(k)$ for intensity level $k \in \{0, \dots, 255\}$ was estimated as

$$h(k) = \frac{1}{N_M} \sum_{\mathbf{r}: M(\mathbf{r})=1} \mathbf{1}[g(\mathbf{r}) = k],$$

where $N_M$ is the number of tissue pixels and $\mathbf{1}[\cdot]$ is the indicator function. The entropy was then defined as

$$H = -\sum_{k: h(k)>0} h(k) \log_2 h(k).$$

This quantity increases with greater diversity of intensity values and serves as a scale-agnostic descriptor of local disorder. This entropy does not assume any specific tissue model and is sensitive only to the distribution of intensities within each tile.

**Edge density and fine-scale microstructural fragmentation.** Microstructural fragmentation was quantified using an edge density measure derived from gradient-based edge detection. The grayscale image $G(\mathbf{r})$ was processed using a Canny edge detector with fixed lower and upper thresholds. This yields a binary edge map $E(\mathbf{r}) \in \{0,1\}$. Edge density was defined as

$$\rho_E = \frac{1}{N_M} \sum_{\mathbf{r}: M(\mathbf{r})=1} E(\mathbf{r}),$$

representing the fraction of tissue pixels identified as part of an edge. Higher values of $\rho_E$ correspond to more fragmented or irregular microstructure. Since the same detector parameters were used across all tiles, differences in $\rho_E$ reflect architectural variation rather than parameter tuning.

**Orientation isotropy and gradient entropy**. To quantify directional coherence and loss of directional organization, we computed an isotropy measure based on the distribution of gradient orientations. Horizontal and vertical gradients were computed using Sobel operators (Lu et al. 2023; Tang et al. 2023; Wang et al. 2024), yielding gradient components $G_x(\mathbf{r})$ and $G_y(\mathbf{r})$. For tissue pixels, gradient magnitude and orientation were defined as

$$m(\mathbf{r}) = \sqrt{G_x(\mathbf{r})^2 + G_y(\mathbf{r})^2}, \theta(\mathbf{r}) = \arctan 2(G_y(\mathbf{r}), G_x(\mathbf{r})) \bmod \pi.$$

Only pixels with $m(\mathbf{r})$ above a fixed percentile threshold were retained to focus on structurally meaningful gradients. The orientation histogram was constructed over $B = 8$ bins on $[0, \pi)$, yielding probabilities $p_i$. Orientation entropy was then computed as

$$H_\theta = -\sum_{i=1}^{B} p_i \log_2 p_i.$$

An isotropy index was defined as the normalized entropy

$$I = \frac{H_\theta}{\log_2 B},$$

with $I = 1$ indicating maximal isotropy and lower values indicating preferred orientations.

**Multiscale entropy change: sensitivity of disorder to coarse graining**. To probe multiscale structure, we quantified how intensity entropy changes under spatial smoothing. The grayscale image was convolved with a Gaussian kernel of standard deviation $\sigma = 1.8$ pixels, yielding a smoothed image $G_\sigma(\mathbf{r})$. Entropy $H_\sigma$ was computed on $G_\sigma$ using the same procedure as for $H$. The multiscale entropy change was defined as

$$\Delta H = H - H_\sigma.$$

Larger values of $\Delta H$ indicate the presence of fine-scale variability that is suppressed by coarse-graining, whereas small values indicate scale-invariant or already smooth structure.

**Construction of a composite disorder measure**. The four quantities $H$, $\rho_E$, $I$ and $\Delta H$ were combined into a single hyperdisorder index. For each feature $X$, a standardized score was computed as

$$Z_X = \frac{X - \mu_X}{\sigma_X},$$

where $\mu_X$ and $\sigma_X$ are the mean and standard deviation across all retained tiles. The hyperdisorder index for tile $j$ was then defined as

$$\mathcal{H}_j = Z_H + Z_{\rho_E} + Z_I + Z_{\Delta H}.$$

This construction assigns equal weight to each disorder mode and yields a dimensionless quantity centered near zero.

**Spatial mapping and patient-level aggregation**. Tile coordinates encoded in filenames were used to reconstruct approximate spatial maps of hyperdisorder within slides. For each slide, $\mathcal{H}_j$ values were plotted at their corresponding $(x, y)$ positions to visualize spatial nonstationarity. Patient-level summaries were obtained by grouping tiles by patient identifier and computing descriptive statistics, including the median hyperdisorder and interquartile range. These summaries were used to characterize between-patient variability without imposing parametric assumptions.

**Software and computational tools**. All computations were performed using Python. Image processing relied on OpenCV, numerical operations on NumPy, data handling on pandas and visualization on Matplotlib. No machine learning models or pretrained networks were used. All parameters were fixed a priori and applied uniformly across samples.

Overall, we have defined the data sources, preprocessing steps, mathematical constructs and computational procedures used to quantify hyperdisorder in gastric cancer histology. The resulting framework is fully operational, reproducible and not specific to gastric cancer, providing a general approach for probing multiscale architectural incoherence in tumor tissues and for systematic comparison across datasets and cancer types.

RESULTS

We report the quantitative results obtained by applying our approach to gastric cancer histology, focusing on distributions, correlations, spatial patterns and patient-level aggregation.

**Tile-level hyperdisorder distribution**. We first quantified hyperdisorder at the tile level across 899 tumor tiles derived from 37 patients. The resulting hyperdisorder index exhibited a broad, asymmetric distribution (Fig. B), spanning from approximately −9.7 to 4.5 in standardized units, with a median close to zero by construction. This wide range indicates that architectural disorder varies substantially between spatially distinct tumor regions, even within the same cohort. Examination of the individual components revealed nontrivial relationships between disorder modes. Texture entropy and edge density were strongly positively correlated (Pearson correlation coefficient $r \approx 0.82$), indicating that increased intensity variability is often accompanied by increased microstructural fragmentation (Fig. C). In contrast, isotropy showed weaker and negative correlations with entropy ($r \approx -0.17$) and edge density ($r \approx -0.07$), suggesting that loss of directional organization is only partially aligned with fine-scale disorder. Multiscale entropy change was negatively correlated with entropy ($r \approx -0.51$) and weakly correlated with edge density ($r \approx -0.22$), reflecting distinct sensitivity to coarse-graining. The composite hyperdisorder index correlated most strongly with edge density ($r \approx 0.76$) and entropy ($r \approx 0.57$), while retaining moderate associations with isotropy ($r \approx 0.46$) and multiscale entropy change ($r \approx 0.22$).

These quantitative relationships show that the hyperdisorder index integrates partially decoupled components rather than collapsing onto a single dominant feature, establishing a multicomponent description of architectural disorder that motivates aggregation beyond individual tiles.

**Patient-level aggregation**. Building on tile-level measurements, we aggregated hyperdisorder values at the patient and slide levels to assess larger-scale organization. Patient-level median hyperdisorder values varied systematically across the cohort, ranging from approximately −3.6 to 2.5 (Fig. D). This spread indicates that tumors differ not only in local variability but also in their overall architectural regime. Within-patient variability, quantified by the interquartile range of tile-level hyperdisorder, remained substantial across most patients, with values typically between 1.0 and 2.5 (Fig. F). Importantly, higher median hyperdisorder did not correspond to reduced within-patient spread, indicating that increased disorder does not stabilize under aggregation. Spatial mapping of hyperdisorder within individual slides further revealed pronounced nonstationarity (Fig. E). In a representative case with 48 sampled tiles, regions of elevated and reduced hyperdisorder appeared interspersed across the tissue section, without a smooth gradient or uniform pattern. These spatial fluctuations were comparable in magnitude to between-patient differences, underscoring that intra-tumoral variability contributes substantially to observed disorder statistics.

Taken together, patient-level summaries and spatial maps show that architectural disorder is unevenly distributed both within and between tumors, reinforcing the need to interpret tumor architecture as a spatially structured, multiscale phenomenon rather than as a homogeneous property.

Our results show that gastric cancer histology displays pronounced tile-level variability in hyperdisorder, partial decoupling among disorder components and systematic differences in patient-level medians, together with strong spatial nonstationarity within individual slides. Taken together, these findings characterize tumor architecture as intrinsically multiscale and unevenly organized, rather than as a uniformly degraded or purely random structure.

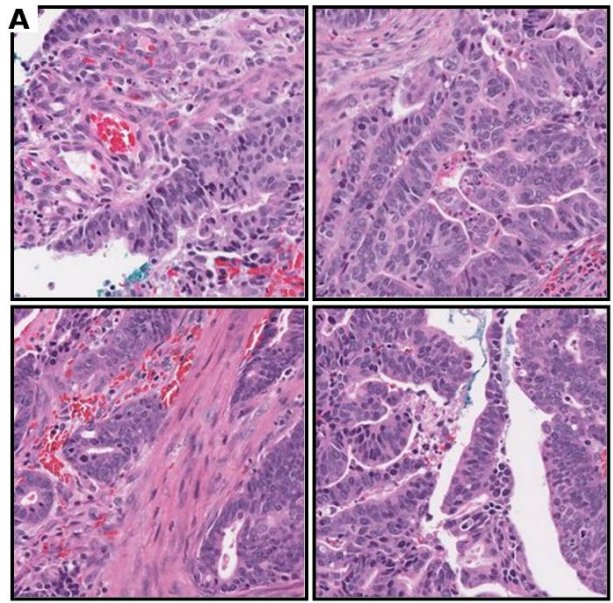
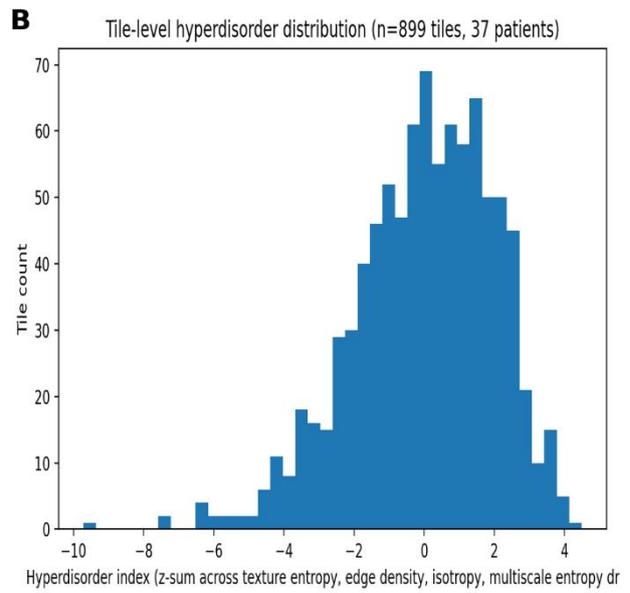
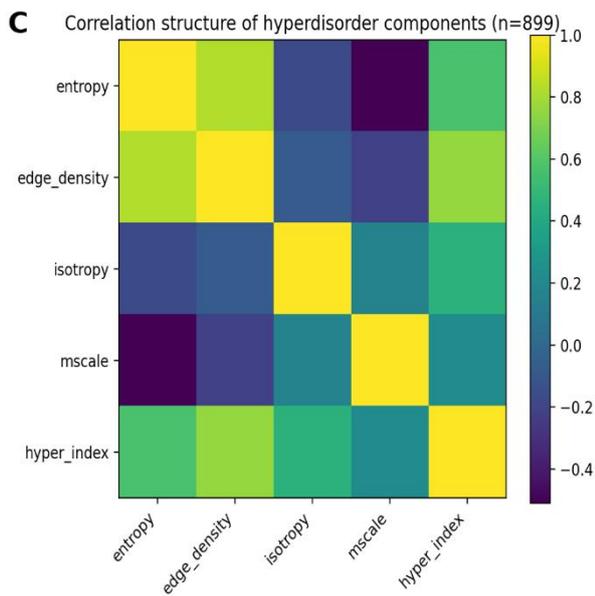
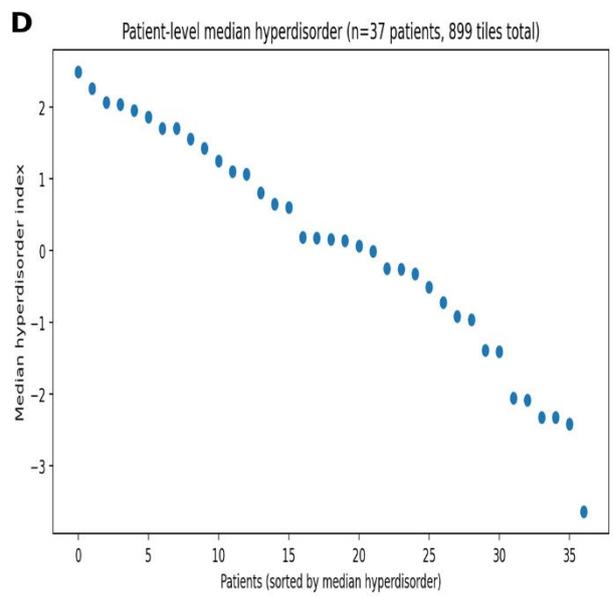
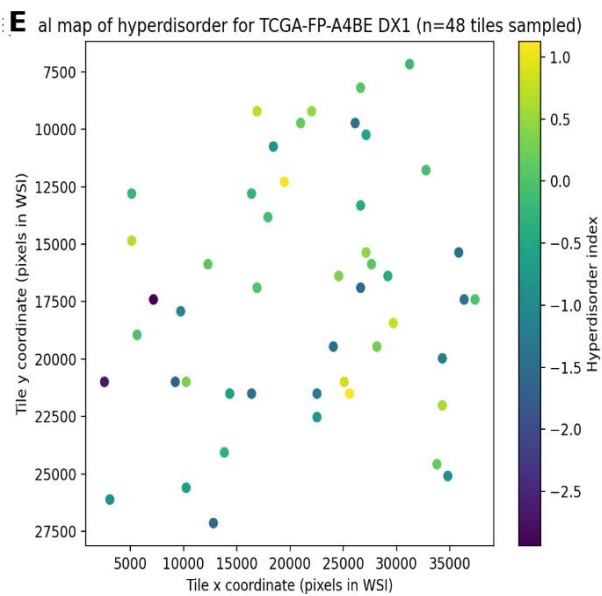
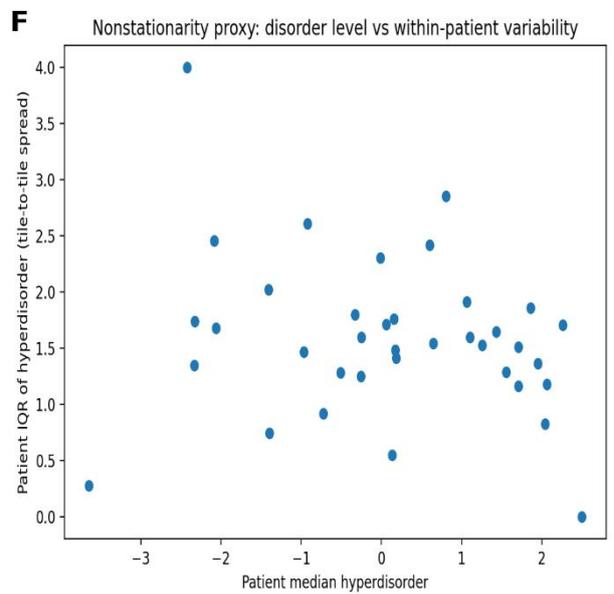

**Figure**. Hyperdisorder as a multiscale architectural regime in gastric cancer histology.
**(A)** Representative hematoxylin and eosin–stained histological tiles from a gastric adenocarcinoma whole-slide image (TCGA-3M-AB46, slide DX1). The four panels show distinct tumor regions sampled from the same specimen at comparable magnification, illustrating pronounced intra-tumoral architectural variability, including differences in cellular density, nuclear arrangement, stromal organization, vascular features and glandular distortion. These images are shown without preprocessing other than resizing and provide context for the quantitative analyses shown in panels B–F.
**(B)** Distribution of the tile-level hyperdisorder index computed across 899 tumor tiles from 37 patients. The index is defined as the sum of standardized contributions from texture entropy, edge density orientation isotropy and multiscale entropy change. The broad and asymmetric distribution indicates substantial variability in architectural disorder across tumor regions.
**(C)** Correlation structure among the individual hyperdisorder components and the composite index. Strong coupling between some measures like entropy and edge density, coexists with weaker or inverse correlations involving isotropy and multiscale entropy change, indicating partial decoupling between disorder modes rather than a single dominant source of variability.
**(D)** Patient-level median hyperdisorder values, with patients ordered by their median score. This representation highlights systematic differences in overall architectural disruption between tumors, beyond local tile-level fluctuations.
**(E)** Spatial map of hyperdisorder values for a representative slide (TCGA-FP-A4BE, DX1), with each point corresponding to a sampled tile plotted at its approximate coordinate within the whole-slide image. The patchy distribution of high and low values reveals strong spatial nonstationarity of architectural disorder within a single tumor section.
**(F)** Relationship between patient-level median hyperdisorder and within-patient variability, quantified as the interquartile range of tile-level values. The persistence of substantial variability across a wide range of median disorder levels indicates that architectural heterogeneity does not converge under spatial aggregation.
Together, panels A–F show that gastric cancer histology is characterized by multiscale architectural disruption, partial decoupling between disorder measures and pronounced spatial nonstationarity, consistent with a hyperdisordered architectural regime rather than uniform or purely random heterogeneity.

QUANTITATIVE PREDICTIONS OF HYPERDISORDER DRIVEN BY TUMOR GROWTH

Calling tumor growth hyperdisordered is not intended as a descriptive label but as a quantitative claim about how heterogeneity behaves under spatial aggregation and how disorder scales with system size. Hyperdisorder implies that architectural variability does not obey the convergence properties expected under ordinary stochastic heterogeneity. As a result, it yields explicit, testable predictions about sampling, scaling and spatial organization that can be assessed using existing tumor datasets. These predictions concern not only the magnitude of heterogeneity but its scaling laws, spatial coupling and dependence on growth geometry.

A first consequence of hyperdisorder is the failure of averaging. If a tumor occupies a hyperdisordered regime, increasing the sampled volume does not necessarily stabilize estimates of genetic, phenotypic or microenvironmental quantities. Let $X(\mathbf{r})$ denote a spatially defined biomarker field like mutation burden, expression level, hypoxia score or immune infiltration. For a sampling window of linear size $R$, define the regional mean
$$\bar{X}_R = \frac{1}{|W_R|} \int_{W_R} X(\mathbf{r}) \, d\mathbf{r},$$
where $W_R$ is a window of radius $R$. Under ordinary randomness with short-range correlations, the variance scales as
$$\text{Var}(\bar{X}_R) \propto R^{-d},$$
with $d$ the effective spatial dimension. Hyperdisorder predicts instead
$$\text{Var}(\bar{X}_R) \propto R^{-\alpha}, \alpha \ll d,$$
or even scale-dependent $\alpha(R)$, implying slow or non-monotonic decay. This prediction is testable using spatial transcriptomics, multiplex immunofluorescence or tiled whole-slide images by fitting $\alpha$ across scales. Falsification occurs if variance converges rapidly beyond a modest $R$.

A second prediction concerns genetic divergence across space. In multiregion sequencing, let $D_{ij}$ denote the genetic distance between regions $i$ and $j$, defined for example by Jaccard distance of mutation sets or by phylogenetic branch length and let $r_{ij}$ denote their spatial separation. Under simple spatial mixing, one expects a stable relationship $D(r)$ with decreasing dispersion as tumor size increases. Hyperdisorder predicts the opposite: as tumors grow, the conditional variance $\text{Var}(D_{ij} \mid r_{ij})$ increases and a single distance-to-divergence curve ceases to be predictive. This can be tested by stratifying published multiregion Whole-Exome Sequencing or Whole-Genome Sequencing cohorts by tumor size or volume proxy. Falsification occurs if $D(r)$ tightens with increasing size.

A third prediction concerns the spatial organization of phenotypic patches. Define patches by thresholding a continuous field $X(\mathbf{r})$ into regions where $X$ exceeds a biologically meaningful cutoff and let $A$ denote patch area. Under scale-aligned heterogeneity, the distribution $P(A)$ is expected to be narrow or exponentially bounded. Hyperdisorder predicts a broad, heavy-tailed distribution,

$$P(A) \sim A^{-\tau},$$

with the tail exponent $\tau$ decreasing as tumor size increases, reflecting growth-amplified patchiness. This prediction can be tested using digital pathology segmentations or radiological feature maps. Falsification occurs if $P(A)$ remains narrowly distributed and scale-invariant.

A fourth prediction concerns coarse-graining and phenotypic classification. Consider a tumor phenotype index $Y_R$ obtained by smoothing or aggregating data at scale $R$ (for example cell scale, 200 μm, 1 mm, 5 mm). Under ordinary heterogeneity, classifications based on $Y_R$ converge as $R$ increases. Hyperdisorder predicts persistent disagreement between classifications at different $R$, quantified for instance by non-vanishing disagreement probability

$$\Pr(Y_{R_1} \neq Y_{R_2}) \not\to 0 \text{ as } R_2 \to \infty.$$

This is testable using spatial omics or multiplexed imaging. Falsification occurs if classifications stabilize at large scales.

A fifth prediction links hyperdisorder explicitly to growth geometry. If hyperdisorder is generated by growth, scaling exponents and patch statistics should differ between spatial regions associated with distinct growth dynamics. Let $\alpha_{\text{edge}}$, $\alpha_{\text{mid}}$, $\alpha_{\text{core}}$ denote scaling exponents measured separately at the invasive front, intermediate zone and necrotic or fibrotic core. Hyperdisorder predicts systematic differences and crossover scales separating these regimes. This can be tested by region annotation in whole-slide images or spatial transcriptomics. Falsification occurs if scaling behavior is spatially homogeneous.

Additional quantitative consequences follow directly from these predictions.
First, the effective number of independent samples $N_{\text{eff}}(R)$ within a region of size $R$ is reduced relative to geometric expectations, scaling as $N_{\text{eff}} \sim R^{\alpha}$ rather than $R^d$.
Second, estimates of extreme values, such as maximum hypoxia or maximum clonal divergence, are predicted to grow logarithmically or algebraically with sampled area rather than saturating.
Third, correlations between distinct biomarker fields $X(\mathbf{r})$ and $Y(\mathbf{r})$ are predicted to be scale-dependent, with cross-correlations $C_{XY}(R)$ changing sign or magnitude as a function of $R$, reflecting partial decoupling of disorder modes.
Minimal datasets are sufficient to test several of these predictions. For example, whole-slide images with basic immunohistochemistry can be tiled at multiple resolutions, per-tile features computed and variance scaling and coarse-graining behavior evaluated directly, already addressing the first and fourth predictions. More detailed datasets could enable simultaneous testing of multiple predictions.

In summary, treating tumor growth as hyperdisordered becomes meaningful only when formulated as a set of quantitative claims about how heterogeneity scales with spatial extent and growth geometry. These claims generate explicit mathematical predictions, can be tested with existing data and admit clear falsification criteria.

CONCLUSIONS

We find that gastric cancer histology exhibits architectural variability that is not adequately captured by single-scale descriptors or by assumptions of uniform heterogeneity. We quantified disorder both at the tile level and aggregated across spatial and patient scales, revealing broad, asymmetric distributions of architectural disruption and partial decoupling between disorder components. Correlation analysis demonstrated that texture entropy and microstructural fragmentation tend to covary, while orientation isotropy and multiscale entropy change behave more independently, indicating that no single feature dominates the architectural signal. Patient-level aggregation further showed systematic differences in median disorder alongside persistent within-patient variability and spatial mapping highlighted pronounced nonstationarity within individual slides. Together, these results indicate that architectural disorder in gastric tumors is unevenly distributed across space and scale and that increased aggregation does not necessarily stabilize estimates. The convergence of these observations supports the interpretation of tumor architecture as a multiscale, spatially structured phenomenon rather than a homogeneous or purely random one.

We shift the description of tumor architecture from local heterogeneity toward a regime-based characterization grounded in scaling behavior. Unlike traditional histopathological grading or single-feature image metrics (McAddy et al. 2020; Tollemar et al. 2020; Lisievici et al. 2021), our approach does not aim to classify tissues by predefined patterns or labels. Instead, it quantifies how disorder is organized across scales and how different disorder modes interact. This distinguishes it from texture-only analyses, which summarize local variability without addressing spatial coupling (Kunimatsu et al. 2022; Mehta et al. 2023; Varghese et al. 2023) and from segmentation-based approaches, which rely on explicit identification of cellular or glandular units (Sun et al. 2022; Ahamed et al. 2023; Sun et al. 2023; Xue, Yao and Teng 2024). Compared with machine learning classifiers, our approach remains interpretable, with each component defined mathematically and linked to a specific aspect of tissue organization (Huang et al. 2018; Antonelli et al. 2019; Alkhathlan and Saudagar 2022; Ghobadi, Emamzadeh and Afsaneh 2022; Ellrott et al. 2025). We treat architectural disorder as a composite, multiscale property whose defining feature is partial decoupling rather than magnitude alone. In this sense, our approach complements rather than replaces existing techniques: it does not compete with molecular profiling, spatial transcriptomics or deep learning–based classification, but provides a quantitative layer that can be integrated with them. By focusing on scaling relations and spatial nonstationarity, we provide a different lens through which architectural complexity can be compared across tumors and datasets, emphasizing structural organization rather than categorical assignment. This perspective situates architectural analysis closer to physical descriptions of disordered systems while remaining grounded in observable histological features, clarifying its position relative to established quantitative pathology methods. Within the landscape of existing methodologies, our approach can be classified as a mesoscopic, scale-aware architectural analysis. It sits between purely local feature extraction and fully global classification schemes and between qualitative histopathology and high-dimensional molecular profiling.

Several limitations must be acknowledged. Our analyses were performed on a subset of available tiles rather than the full dataset, which constrains statistical power and limits generalization. Reported correlations and distributions are descriptive and cohort-specific and scaling predictions remain theoretical within the present data. No normal gastric tissue was included, preventing direct comparison between tumor and non-tumor architectural regimes. Additionally, the hyperdisorder index assigns equal weight to its components, a choice that is operational rather than derived from optimization or biological priors. Spatial coordinates were approximated from tile metadata rather than reconstructed from full whole-slide geometry, which may blur fine spatial relationships. Finally, while mathematical formulations are explicit, some predictions require datasets with richer spatial resolution or longitudinal sampling to be tested rigorously.

Potential applications of our approach arise from its ability to describe how tumor architecture changes with spatial scale and growth. In research settings, it provides a basis for comparing tumors in terms of architectural regimes rather than isolated features, thereby supporting cross-cohort and cross-modality analyses. Because our approach operates directly on spatially resolved data, it is applicable to whole-slide histology, spatial transcriptomics, radiomics and multiregion sequencing datasets, provided spatial coordinates are available. From a diagnostic perspective, our framework provides a means to assess when architectural estimates are stable under aggregation and when they are intrinsically scale-dependent, indicating whether biopsy-based measurements can be reliably generalized to the whole tumor or are dominated by local sampling effects. In clinical contexts, it supports the analysis of sampling strategies by identifying when additional sampling improves reliability and when it fails to do so because architectural variability persists across scales. From a therapeutic standpoint, focusing on scaling behavior allows systematic assessment of how tumor growth geometry and spatial organization influence treatment response and resistance, without requiring assumptions about the underlying mechanisms. Future work may extend our approach to normal tissues, additional cancer types and longitudinal samples and further refine the hyperdisorder index by incorporating alternative weighting schemes or additional modes of disorder.

In conclusion, we asked whether tumor growth can be meaningfully described in terms of a multiscale architectural regime rather than as a random accumulation of local heterogeneity. We found that gastric cancer histology exhibits spatially structured disorder characterized by partial decoupling across scales, persistent nonstationarity and limited stabilization under aggregation. This suggests that tumor architectural complexity is best understood through its scaling behavior rather than isolated local descriptors, enabling quantitative comparison and explicit falsification.

**DECLARATIONS**

**Ethics approval and consent to participate.** This research does not contain any studies with human participants or animals performed by the Author.
**Consent for publication.** The Author transfers all copyright ownership, in the event the work is published. The undersigned author warrants that the article is original, does not infringe on any copyright or other proprietary right of any third part, is not under consideration by another journal and has not been previously published.
**Availability of data and materials.** All data and materials generated or analyzed during this study are included in the manuscript. The Author had full access to all the data in the study and took responsibility for the integrity of the data and the accuracy of the data analysis.
**Competing interests.** The Author does not have any known or potential conflict of interest including any financial, personal or other relationships with other people or organizations within three years of beginning the submitted work that could inappropriately influence or be perceived to influence their work.
**Funding.** This research did not receive any specific grant from funding agencies in the public, commercial or not-for-profit sectors.

**Acknowledgements:** none.
**Authors' contributions.** The Author performed: study concept and design, acquisition of data, analysis and interpretation of data, drafting of the manuscript, critical revision of the manuscript for important intellectual content, statistical analysis, obtained funding, administrative, technical and material support, study supervision.
**Declaration of generative AI and AI-assisted technologies in the writing process.** During the preparation of this work, the author used ChatGPT 5.2 to assist with data analysis and manuscript drafting and to improve spelling, grammar and general editing. After using this tool, the author reviewed and edited the content as needed, taking full responsibility for the content of the publication.